# Development of the TIP-HOLE gas avalanche structure for nuclear physics/astrophysics applications with radioactive isotope beams: preliminary results.

**Jaspreet Singh Randhawa, Marco Cortesi, Wolfgang Mittig, Thomas Wierzbicki, Alejandro Gomez**

National Superconducting Cyclotron Laboratory (NSCL)
Michigan State University (MSU)
East Lansing, Michigan 48824, U.S.A.

Cortesi@nscl.msu.edu

**Abstract**. We discuss the operational principle and performance of new micro-pattern gaseous detectors based on the multi-layer Thick Gaseous Electron Multiplier (M-THGEM) concept coupled to a needle-like anode. The new gas avalanche structure aims at high-gain operation in nuclear physics and nuclear astrophysics applications with radioactive isotope beams. It is thereafter named TIP-HOLE gas amplifier, and consists of a THGEM or a two-layers M-THGEM mounted in a WELL configuration. The avalanche electrodes are collected by thin conductive needles (with up to a few ten um radius and a height of 100 um), located at the center of the hole and acting as point-like anode. The bottom area of the needle may be surrounded by a cylindrical cathode strip in order to increase the electron collection efficiency. The electric field lines from the drift region above the M-THGEM are focused into the holes, and then forced to converge on the needle tip. An extremely high field is reached at the top of the needle, creating a point-like avalanche process. Stable, high-gain operations in a wide range of pressures may be achieved at relatively low operational voltage, even in pure quencher gas at atmospheric pressure (e.g. pure isobutene). The TIP-HOLE structure may be produced by the innovative scalable additive manufacturing technology for large-area, multiple-layer printed circuit boards, recently developed by the UHV technology company (USA) and discussed for the first time in this work.

## 1. Introduction

At current and future radioactive beam facilities (e.g. NSCL and FRIB [1], and others), shortest-lived rare isotopes will be available at a large range of energy (from a few MeV/u up to a few hundred MeV/u) and their properties, such as excitation level schemes, can be studied via nuclear reactions in a collision with a stable target. Many reaction channels are typically open and an event-by-even identification of the reaction products in the exit channel is required.

Innovative detector designs are emerging to study the nuclear reactions with low intensity beams, including Time Projection Chamber operated in Active-Target mode (AT-TPC) [2]–[5]. The design of innovative Micro-Pattern Gaseous Detectors (MPGDs) as position-sensitive readouts for TPC and specifically dedicated to heavy-ion tracking, aims at finding an alternative structure to standard gas avalanche schemes to fulfil the most stringent constraints imposed by future experiments with rare isotope beams. This includes operation in harsh experimental conditions at an ever increased demand for higher detection performance. An instructive example is the large class of inverse-kinematic reaction studies that require simultaneous particles tracking in an active-target gas medium (active target mode), and precise measurement of the kinetic energies and escaping angles of the reaction products [6]. The latter are needed to reconstruct all the kinematic variables of the reaction, as well as for an accurate particle identification. When no magnetic field is used, the above tasks can be attained only by stopping completely the charged reaction products inside the detector effective volume, using a high stopping power gas medium - for instance isobutene as a proton target, or deuterated isobutene as a deuterium target.

However, it is extremely challenging to achieve a stable high-gain operation in complex molecules or in pure gas quenchers such as isobutene. In fact, complex molecules are characterized by many vibrational/rotational modes that can be excited by electron impacts, in competition with the ionization process, up to relative high electron energy. In consequence of that, electron multiplication can be achieved only at high electric fields, leading to a high probability of sporadic discharges and a lower stability.

An extremely high electric field strength can be attained by converging the field lines in a single point (dot) or along a thin conductive line (needle). Early attempts to create micro-needle amplification structures were made as early as 1976 by Spindt *et al.* [7], followed by others. In micrometer needle-like multiplier the electric field rapidly decreases as the distance from the needle increases, so that the avalanche electrons do not have the sufficient number of mean free path available to create a substantial multiplication, particularly in standard gas mixtures (i.e. Ar-based mixtures) and at normal (atmospheric) pressure. As a result, no significant gas gain is achievable.

In the later '80s, implementation of the photolithography technique and the silicon-wafer manufacturing technology open new possibilities for alternative needle-like designs, leading to a series of more successful architectures. An example is the Microdot detector (μDOT) introduced by Biagi *et al.* [8] and produced by microelectronic technology, consisting of high granularity anode dots surrounded by a cathode rings, with a pitch of a few tens of micro-meters. In line with the same three-dimensional avalanche configuration, other designs were developed configuring a central thin anode surrounded by a cathode, including the Micro-Pin Array (MIPA) [9] and Micro-Well [10]. More recently Lombardi *et al.* developed a new anode-point based readout named as leak microstructure [11], [12], consisting of a bodking needle acting as anode and placed in a hole (Ø = 400 μm) made of a conducting layer (cathode). Energy resolutions of about 8% FWHM was recorded with α-particles from a 241-Am source in pure isobutene at atmospheric pressure. In principle, all the above structures have a common operational principle and similar properties, characterized by modest gas gain provided by a single multiplication stage, in an open-geometry.

In this work, we present for the first time an alternative MPGD, the TIP-HOLE multiplier, which combines different detector concepts (multi-layer hole-type geometries, patterned surfaces, and needle-type anode) in single robust micro-structures. The aim of the new gas avalanche multiplier is to achieve a stable high gain operation in high stopping power gas mixtures in a wide range of pressures, with the aims of high charge particle tracking capability for nuclear physics applications with rare isotope beams.

## 2. The TIP-HOLE concept.

The TIP-HOLE multiplier concept consists of a THGEM-based structure [13], [14], either one or two-layer M-THGEM mounted in a WELL configuration [15]. The readout consists of a thin needle at the bottom of the holes, which serves as an anode by collecting the avalanche electrons (Figure 1a). Each insulator substrate of the M-THGEM is 400 μm thick, while the diameter and the pitch of the holes are respectively of 0.5 mm and 1 mm. The inner and outer copper surfaces have an etched rim of 0.1 mm around the hole in order to reduce the probability of spurious discharged. At the bottom of the multi-layer THGEM holes, the anode consists of a conical needle of 100 μm height. The spherical tip of the needles has a radius of 10 μm while the base has a radius of 25 μm. Optionally, the needle can also be surrounded by a cathode ring (negatively biased with respect to the anode), permitting to have a better focusing of the avalanche processed towards the needle tip. Neighbor needles may be patterned in either a square or hexagonal close packed lattice, or be interconnected in strips to perform position capability.

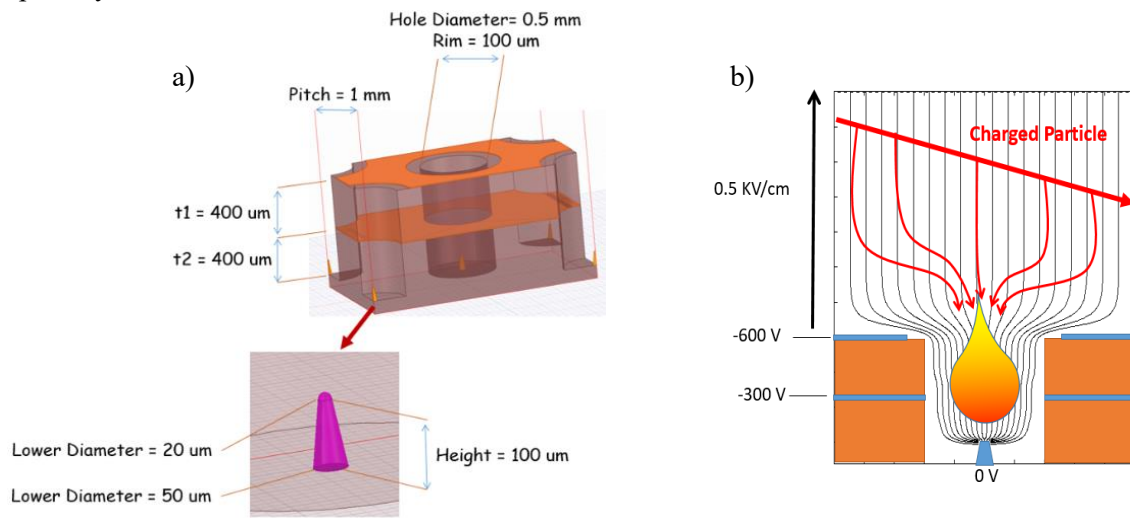

*Figure 1. Schematic drawing depicting typical dimensions and geometry (part a) and operational principle (part b) of the TIP-HOLE detector.*

Ionization electrons, produced anywhere in the gas volume between the M-THGEM top surface and the cathode electrode, are drifted towards M-THGEM holes and focused into the hole by the strong dipole electric field (Figure 1b). If the detector is filled with high-field operation gas mixtures (i.e. isobutene), in which high gains require operation at very high electric field, the M-THGEM will only serve to converge the electrons towards the needle where they will experience an avalanche multiplication in the gas close to the surface of the wire-point anode. For low-pressure operation or in standard gas mixtures (Ar-based mixtures), characterized by long electron mean free paths, the avalanche process will occurs within the M-THGEM holes and the needle anode with serve only to collect the avalanche charge for signal readout. As a result, a large variety of gas mixtures and a wide range of pressures can be used with a single readout gas avalanche readout. Because of the confinement of the avalanche either within the M-THGEM holes, or in the proximity of the needle head, lower photon-mediated secondary effects are expected, as well as a good stability.

Figure 2 illustrate the map of the electric field strengths inside the M-THGEM holes for two proposed needle readout configurations. The field maps were computed using Ansys Electronics Suite (Ver. 19.0.0). In the most simple scheme, the needles are mounted directly on a flat metal board that serves as an anode (at ground), resulting in a strong electric field only on the proximity of the needle tip. The field map depicted Figure 2 was computed assuming a voltage difference of 400 Volt applied symmetrically to the top and bottom M-THGEM multiplication stages. Because of the electrons diffusion, a large fraction of the electrons may be collected on the flat boards, far away from the needle, being subjected to a low electric field. This, in turn, corresponds to a low an overall gas gain and potentially a loss in energy resolution – the multiplication factor will depends on the trajectory of the electron inside the M-THGEM dipole-like field as well as where the electron will be collected on the anode electrode.

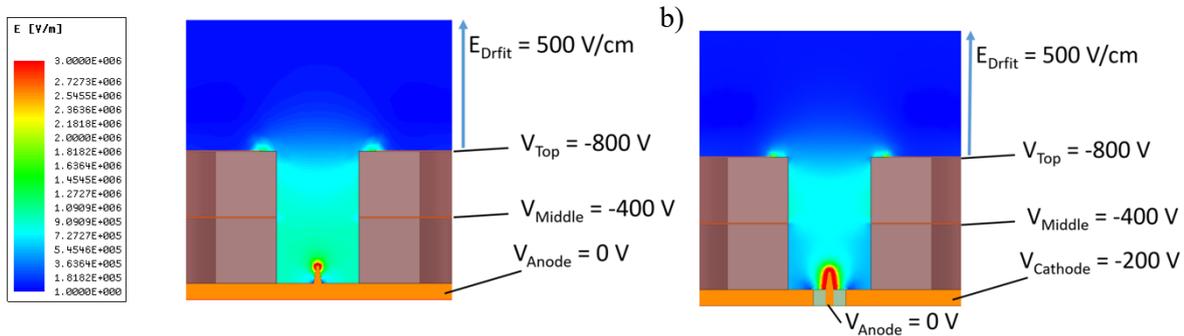

*Figure 2. Electric field strength computed using Ansys Electromagnetic Suite (Ver. 19.0.0), for the two proposed detector configurations: the needles are either connected to a uniformly biased bottom electrode (part a) or surrounded by a cathode (negatively biased) ring (part b).*

A potential better gas gain, as well as a better avalanche electrons collection and energy resolution, may be achieved by electrically isolating the needles for the WELL bottom readout board by a cathode ring, negatively biased with respect to the needle (at -200 Volt in Figure 2b). This imposes the electric field lines to fully converge on the needle resulting on a high electric field along the whole needle length. However, this technical solution has some expected drawbacks, including a possible charging-up of the insolation strips between the needle and the cathode ring, with effects on long-term gain stability, a more complex operation, and a more expensive manufacturing procedure.

## 3. The prototype and preliminary results

We have carried out a preliminary performance evaluation on a fist TIP-HOLE detector prototype, fabricated using a similar technique proposed by Lombardi et al. and described in ref. [12]. The detector consists of two independent PCB boards. The bottom board serves as an anode and comprises a set of 300 μm diameter holes arranged in a 5x5 matrix (Figure 3a). Tattoo needles were made of nickel and have a with a maximum radius of 350 μm and a round tip of 30 μm radius (Figure 3b). They were inserted in the holes and commonly connected to the back layer of the PCB for extracting the output signals. A distance of 1 mm separates one needle from another. The height of the needles (in the range of 0-500 μm) was precisely controlled by means of micro-meter spacers, with a tolerance of 10-20 μm.

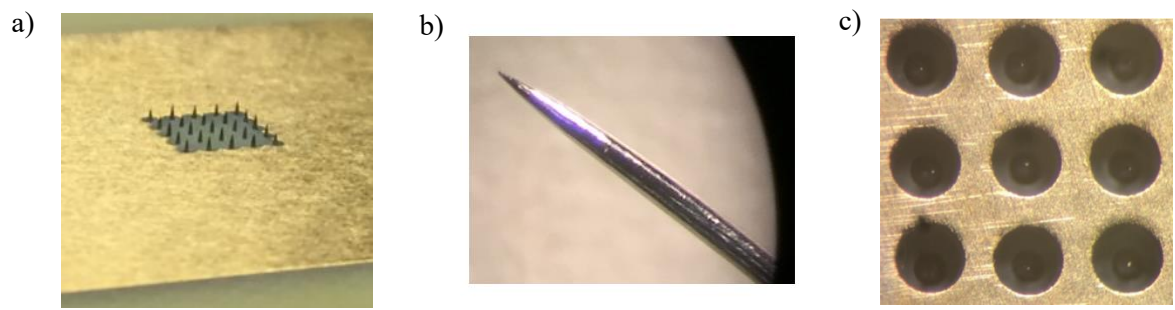

*Figure 3. Part a) PCb boards used as based readout for the TIP-HOLE prototype, including 25 needles arranged in a 5x5 matrix mounted in the boards. The needles are interconnected to the bottom layer of the PCB for reading out the signals. Part b) Photograph of the Tattoo needles used as anode. Part c) Top-view photographs of the final TIP-HOLE assembling.*

A second, top PCB board, incorporating the two-layer M-THGEM structure, with a 5x5 (500 μm diameter) holes matrix at a pitch of 1 mm, was placed directly on top of the needle board. The alignment of the two PCBs was controlled by means of fiducials and manually adjusted with the help of a microscope. A top-view photograph of the final TIP-HOLE detector is shown in Figure 3c.

The TIP-HOLE detector was completed by an upper grid that function as a cathode, placed 3 cm above the M-THGEM top surface. The detector was filled with P10 (Ar/10%CH$_4$), at different pressures (from 200 Torr to 760 Torr), and operated in continuo gas flow (with a few sccm). A collimated 5.5 MeV alpha-particle source (241-Am) was positioned on top of the cathode grid. Signals were readout from the 25 needles, commonly interconnected, and processed by a charge sensitive preamplifier (Tennelec FET Charge Sensitive Pre-amplifier model TC 175b).

As shown in Figure 4, we found that the shape of the pulses strongly depends on the bias configuration. In a symmetrical bias configuration, where equal voltage difference is symmetrically applied to the first and second M-THGEM stage, the output signals are characterized by two peaks separated by 2 µs (Figure 4b) – the time interval between the two peaks (within a few µs) depends actually on the gas pressure (200 Torr in the case of Figure 4b). This effect was never been observed in the case of conventional multi-layer THGEM equipped with flat PCB readout. The double peaks are believed to be originated by avalanche ions that, created in the proximity of the needles, drift back in a high electric field present on the volume of the top M-THGEM stage, inducing a second late signal on the needle anode according to the Shockley–Ramo theorem [16], [17]. By reducing the field strength on the top M-THGEM stage, such that it acts as a low gain pre-amplification stage, the second peak can be strongly reduced (Figure 4b) without affecting the total detector gas gain – the main gas gain will be provided by the single-point avalanche process developed around the needles.

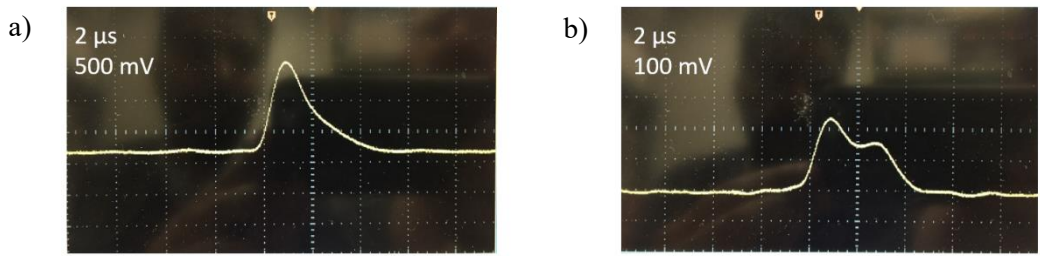

Figure 4. Typical TIP-HOLE output signals, processed by a charge sensitive preamplifiers, resulted from an optimized bias configuration (low pre-amplification gain on the top M-THGEM stage) and a symmetrical bias configuration (high gas gain on the top M-THGEM stage).

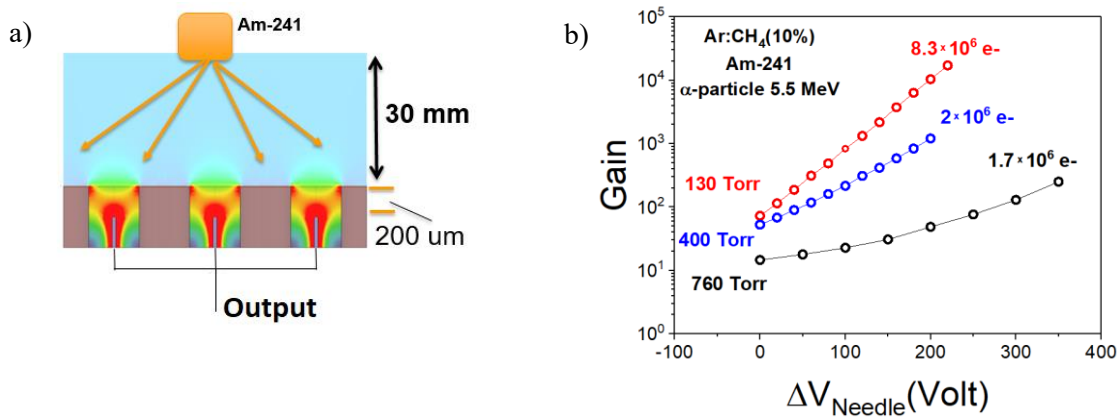

Figure 5. Schematic drawing of the experimental Set-up and gas gain curve measured in P10 at different pressure with a first TIP-HOLE prototype, by irradiating the detector with 5.5 MeV alpha-particles.

Figure 5b shows the gain curve measured in P10 at different pressure, for 5.5 MeV alpha-particle crossing the 30 mm drift region on top of the TIP-HOLE, as depicted by Figure 5a. The experimental setup consists of a conventional single-layer THGEM, with 800 μm thick holes on top of 500 μm high needles - so that the distance between tip of the needles and the top THGEM surface was of 200 μm. The gains were obtained by applying a fixed voltage on the THGEM to surface (-400 Volt) and changing the (positive) voltage bias on the needles. The maximum achievable gain depends on the pressure of the filling gas, reaching a value above $10^4$ in the case of 130 Torr. For atmospheric pressure operation, the maximum achievable gain was limited by the large number of primary deposited by the alpha-particle in the drift region. However, a total avalanche charge of the order of $10^6$ was reached in all the tested pressures, a value which is close to the experimental Reather limit value ($10^6$-$10^7$).

## 4. Future plan: 3D printing fabrication

We have observed that the large ambiguity in setting the height of the needles and their positioning in the center of the holes was the main source of pulse-height fluctuations. In this work, the detector assembling, including needle position and PCB alignment, was performed manually with limited diagnostics possibility. The detector resulted to be affected by a quite poor energy resolution - typically 2-3 time worse resolution compared to a same-geometry two-layer M-THGEM. Better performance in term of gas gain uniformity and large-area detector requires a better automatic fabrication methodology.

An attractive solution may be offered by the modern scalable additive manufacturing technology for a large area, multiple layer printed circuit boards developed by the UHV Technologies industry, fully supported by the USA Department of Energy (grant # DE-SC0017233). The new technique employs confined electro-plating 3D printer to fabricate high conductivity copper features at room temperature. One of the key advantages of this technology is 3-D fabrication of both plastic and metallic features in the same 3-D printer, enabling many innovative micro-pattern gaseous detector architectures. A collection of nanostructure enable by the new manufacturing approach are shown in Figure 6.

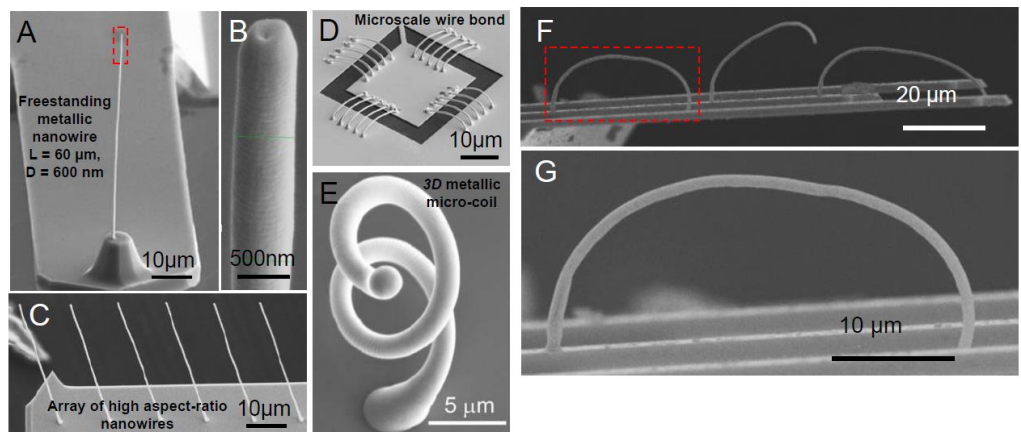

*Figure 6. SEM photographs of prototype nanostructures fabricated using the scalable additive manufacturing technology, the innovative technique proposed for the fabrication of new TIP-HOLE structure. (A) A platinum (Pt) nanowire with an aspect-ratio of 100 (L=60 μm, D=600 nm). (B) Magnified view of the tip of the wire from the box in A. (C) Arrays of cupper (Cu) nanowires with an aspect-ratio of 100. (D) Microscale wire-bond. (E) 3D metallic micro-coil. (F,) and (G) 1 μm diameter Pt 3D spanning interconnects.*

## 5. Conclusions

We have presented and discussed for the first time the operational principle of a new gaseous avalanche structure, the TIP-HOLE detector. The detector consists of a needle anode for charge multiplication and collection embedded in a multi-layer THGEM mounted in a WELL configuration.

In contrast to the previous efforts, the TIP-HOLE detector combined the closed geometry of a multi-layer hole-type architecture (M-THGEM) and needles readout – so the avalanche is developed inside the M-THGEM with lower sensitivity to photon-mediated secondary effect. The single-point avalanche developed in the proximity of needles allows for a fast collection of the electron avalanche. Significant pre-amplification can be achieved in the M-THGEM hole that precedes the needled readout, important for operation at low-pressure or in a gas mixture characterized by a long electron mean free path. The performances, properties and optimization effort of a first prototype have been also reported here. Pulse-height from 5.5 MeV alpha-particle we recorded in P10 gas, showing that high avalanche charge (around $10^6$ electrons per event) can be achieved in various pressure.

**Acknowledgment**
This work supported by the U.S. Department of Energy, Office of Science, Office of Nuclear Physics, under Award Number DE-SC0017233. The authors would like to thank M. Lombardi for helpful correspondence and some provided material from his research on Leak Microstructures.